# Large Scale Artificial Neural Network Training Using Multi-GPUs

Linnan Wang, Wei Wu, Yi Yang, Jianxiong Xiao

## ABSTRACT
This paper describes a method for accelerating large scale Artificial Neural Networks (ANN) training using multi-GPUs by reducing the forward and backward passes to matrix multiplication. We propose an out-of-core multi-GPU matrix multiplication and integrate the algorithm with the ANN training. The experiments demonstrate that our matrix multiplication algorithm achieves linear speedup on multiple inhomogeneous GPUs. The full paper of this project can be found at [1].

## 1. INTRODUCTION
The recent renaissance of the multi-layer Artificial Neural Network (ANN), also referred to as Deep Neural Network (DNN), has demonstrated its effectiveness in a variety of challenging identification tasks such as speech and image recognition [2][3][4]. In general, the number of objects that can be identified by a system ties to the number of neurons in the output layer. Therefore scaling up the size of neuron networks is necessary for increasing the recognition capabilities of the current model.

Training a large scale deep ANN is challenging because the training process not only involves intensive computation but also many training parameters. In this poster, we present a method that can handle both issues by implementing an out-of-core multi-GPUs matrix multiplication. First, we demonstrate the reduction of ANN training to matrix multiplication. Second, we introduce an out-of-core multi-GPUs matrix multiplication with a dynamic task scheduling runtime embedded. The scheduling runtime features a load balancer that offloads tasks according to the devices' real time demand for tasks. It also features a protocol of cache coherence that unifies the multi-GPUs memory spaces to maximize the data reuse. Finally, we integrate the proposed matrix multiplication to the ANN training. The experiments on 4 GPUs demonstrate up to 60x speedup in terms of CPU based training and 2.5x speedup in terms of single GPU training. The out-of-core GPU operations also enable a much larger ANN model size than does the GPU in core training,

## 2. REDUCING THE ANN TRAINING TO MATRIX MULTIPLICATION

During the ANN training, there are two passes before updating the network parameters at each step. The first one is the forward pass, which calculates the network outputs based on a batch of given inputs and the network parameters. The other is the backward pass, which back-propagates the output error and its derivatives to the network parameters of each layer. Equation 1, as shown on the poster, demonstrates that the input of a network layer during the forward pass is the multiplication of the output and the parameter matrices in the prior network layer. Equation 3 and 4 outline the error derivatives W.R.T network parameters and outputs at each layer in the backward pass, which still is matrix multiplication. According to the dimensions specified in equations 1, 2, 3 and 4, increasing the number of neurons or training batch increases the scale of matrix multiplication. Therefore, a fast matrix multiplication algorithm is essential for large scale ANN training.

## 3. THE MULTI-GPU MATRIX MULTIPLICATION ALGORITHM
### 3.1 Tile Algorithm
Tile algorithm logically partitions a matrix into its tiled representation. Given tile size = $T$ in a matrix of size $N \times N$, it creates $\lfloor N/T \rfloor * \lfloor N/T \rfloor$ square tiles of size $T \times T$ and ($\lceil N/T \rceil * \lceil N/T \rceil$) - ($\lfloor N/T \rfloor * \lfloor N/T \rfloor$) non-square tiles. Furthermore, the algorithm treats tiles, uniquely indexed by row and column, as the basic elements in a matrix in lieu of scalars. Operations on matrices are subsequently reduced to operations on tiles. As our focus is matrix multiplication, we assume the tile indices of the output matrix are [$i, j$], and the tile indices of the matrix to the left side of multiply operator are [$i, k$] and the tile indices of matrix to the right of multiply operator are [$k, j$]. Hence tile indices $i$ and $j$ uniquely identify a tile, $C_{ij}$, in the output matrix while the upper bound of $k$ represents the required steps to solve $C_{ij}$.

### 3.2 The Scheduling Runtime
The poster presents the infrastructure of our dynamic task scheduling runtime, which consists of four major components:

- *GPU Computation Thread*: It is a CPU thread to submit tasks for a specific GPU. To avoid the OS scheduling preemption, we bind the thread to a dedicated CPU core. Overlapping the communication and computation on a GPU requires at least 2 tasks concurrently running on the streams. Wei et al. [5] demonstrated no performance gain if adopted streams exceed 4. Therefore, we default 4 concurrent tasks for a GPU with each on a GPU stream.

- *CPU Computation Thread*: It is a CPU thread to submit tasks for the rest of CPU cores. Peng et al. proposed the hybrid tile layout to CPU Cores due to the inherent devices' differences [6]. We adopt the same concept but different approach. The CPU cores dequeue a task and solve the task with a multithreaded BLAS kernel, where the tile size is further factorized.

- *Reservation Station (RS)*: It is a buffer designed to hold the upcoming tasks for the GPU computation threads. The runtime conducts work stealing and priority scheduling on it. Each slot of the RS corresponds to a CUDA stream, the purpose of which is to dispatch tasks in the RS to CUDA streams.

- *Non-blocking Task Queue*: It is a non-blocking queue allowing efficient concurrent dequeue and enqueue operations based on the algorithms proposed by Maged and Michael [7].

## 3.3 The Scheduling Strategy

- *work sharing*: In work sharing, the scheduler attempts to assign more tasks to the underutilized processors. The global task queue simulates the work sharing by distributing tasks with respect to the processors' demands.

- *work stealing*: In work stealing, a underutilized processor takes the initiative to steal work from the overloaded processors. Our runtime enables GPU or CPU to steal tasks from the RS of other GPUs when the global task queue depletes.

## 3.4 The Multi-GPU Cache Coherence

In large-scale matrix multiplication, tiles can flow into the GPU memory spaces multiple times, introducing unnecessary communication cost. Caching tiles on the GPU memory spaces can reduce the communication cost. We implement data reuse based on the concept of cache hierarchies, introducing 2 levels of tile cache hierarchies designated by L1 and L2 tile caches.

L1 tile cache stands for the private memory space on the single GPU. If a tile exists on the L1 tile cache, the GPU can directly compute with that tile. L2 tile cache stands for the entire combined memory spaces of all the GPUs on the machine. Since GPU Peer-to-Peer (P2P) data transfer communicates directly via PCI-E switcher, it is more cost-effective to retrieve a tile from the closest GPU according to the hardware proximity if the tile is absent from L1 tile cache. This is the scenario of a L2 tile cache hit. When a tile does not exist on any GPU, the runtime retrieves the data from the host memory.

## 3.5 The Matrix Multiplication Algorithm on Multi-GPUs

In this section, we present an overview of the proposed multi-GPU matrix multiplication algorithm.

| Algorithm 1. The Algorithm for the proposed MultiGPU matrix multiplication |
|---|
| 1. initialize Task Queue (TQ) |
| 2. initialize Cache Coherence Protocol (CCP) |
| 3. spawn computation threads for each GPU |
| 4. bind each GPU thread to a CPU core |
| 5. While (TQ.size != empty) |
| 6.     For each: |
| 7.         Reservation Station(RS) slot = dequeue from TQ |
| 8.     synchronize all the streams to finish the current tasks |
| 9.     For each RS slot: |
| 10.       tile indices $i, j$ = decode(task id) |
| 11.       For each $k$: |
| 12.         A$[i, k]$ = CCP.L1_L2_hit() |
| 13.         B$[k, j]$ = CCP.L1_L2_hit() |
| 14.         if( A$[i, k]$ == NULL ) Get A$[i, k]$ from host RAM |
| 15.         if( B$[k, j]$ == NULL ) Get B$[k, j]$ from host RAM |
| 16.         Asynchronous GEMM(A$[i, k]$, B$[k, j]$) |
| 17.     Asynchronous harvest C$[i, j]$ from all streams |

## 4. EXPERIMENT RESULTS

Our experiments are conducted on a shared memory machine with 2 NVIDIA K40c and 2 NVIDIA TITAN X GPUs. The experiments are split into two parts: the first part demonstrates the performance of the out-of-core multi-GPUs matrix multiplication algorithm. The second part presents the speedup of a forward and backward pass after integrating the aforementioned algorithm.

## 4.1 The Performance of the Matrix Multiplication in Single Precision

The initial results in the poster demonstrate that our matrix multiplication algorithm can achieve the max speed of 16.8 Teraflops in single precision on the test machine, which is equivalent to the summation of all GPUs' practical peaks. When the matrix size is less than $10^4$, the workload cannot fully saturate the GPU computing ability. Therefore the performance increases along with the matrix size while the size is less than $10^4$ and plateaus afterwards. The profiling data on the poster indicates the cache coherence logic dramatically reduce the redundant data transfer.

## 4.2 ANN Training Integration

We integrate the matrix multiplication algorithm with Caffe [2] to investigate the ANN training performance. Details about network setup can be found in the poster. For the training dataset, we adopt the CIFAR 10 that contains 60000 pieces of labeled 32x32 color images.

We benchmark the performance with the Caffe's built-in benchmark utilities that measure the elapsed time for a forward and backward pass. We sampled 10 passes and took the average. The experiment data demonstrates that we can speed up ANN training up to 2.48 W.R.T Caffe's in-core GPU training and 62.3 W.R.T Caffe's CPU training. In addition, the out of core GPU operation enables us to train a much larger network than the Caffe's GPU in-core training.

## 5. REFERENCES

[1] Wang, L., Wu, W., Xiao, J., & Yang, Y. (2015). BLASX: A High Performance Level-3 BLAS Library for Heterogeneous Multi-GPU Computing. arXiv preprint arXiv:1510.05041.

[2] Jia, Y., Shelhamer, E., Donahue, J., Karayev, S., Long, J., Girshick, R., ... & Darrell, T. (2014, November). Caffe: Convolutional architecture for fast feature embedding. In Proceedings of the ACM International Conference on Multimedia (pp. 675-678). ACM.

[3] LeCun, Y., Bengio, Y., & Hinton, G. (2015). Deep learning. Nature, 521(7553), 436-444.

[4] Schmidhuber, J. (2015). Deep learning in neural networks: An overview. Neural Networks, 61, 85-117.

[5] Wu, W., Bouteiller, A., Bosilca, G., Faverge, M., & Dongarra, J. (2015, January). Hierarchical DAG Scheduling for Hybrid Distributed Systems. In 29th IEEE International Parallel & Distributed Processing Symposium.

[6] Song, F., Tomov, S., & Dongarra, J. (2012, June). Enabling and scaling matrix computations on heterogeneous multi-core and multi-GPU systems. In Proceedings of the 26th ACM international conference on Supercomputing (pp. 365-376). ACM.

[7] Michael, M. M., & Scott, M. L. (1996, May). Simple, fast, and practical non-blocking and blocking concurrent queue algorithms. In Proceedings of the fifteenth annual ACM symposium on Principles of distributed computing (pp. 267-275). AC